# Aerodynamic Drag and Gyroscopic Stability

Elya R. Courtney and Michael W. Courtney

BTG Research, Baton Rouge, Louisiana

Michael_Courtney@alum.mit.edu

**Abstract:** This paper describes the effects on aerodynamic drag of rifle bullets as the gyroscopic stability is lowered from 1.3 to 1.0. It is well known that a bullet can tumble for stability less than 1.0. The Sierra Loading Manuals (4[th] and 5[th] Editions) have previously reported that ballistic coefficient decreases significantly as gyroscopic stability, Sg, is lowered below 1.3. These observations are further confirmed by experiments reported here. Measured ballistic coefficients were compared with gyroscopic stabilities computed using the Miller Twist Rule for nearly solid metal bullets with uniform density and computed using the Courtney-Miller formula for plastic-tipped bullets. The relationship between Sg and drag may be used to test the applicability of existing gyroscopic stability formulas for given bullet designs and to evaluate the accuracy of alternate formulas in cases where the existing stability formulas are not as accurate. The most definitive test of formulas predicting stability will always be observation of whether bullets tumble under given conditions. However, observations of drag changes provide valuable supplemental information because they suggest changes in stability as conditions change. Use of a continuous variable (drag) rather than a binary variable (tumbling) allows insight into stability over a range of conditions where the binary variable does not change.

## Introduction

Prior to publication of the Miller Twist Rule (Miller, 2005), the most common methods for estimating bullet stability were the relatively inaccurate Greenhill formula and the expensive computer modeling program called PRODAS. The modeling program is known to be accurate and is widely used by government laboratories and bullet companies. Bullet companies usually publish a minimum recommended twist rate, usually a twist that will ensure bullet stability under the most dense atmospheric conditions a hunter or recreational shooter is likely to encounter, with some margin of error so that no shooter is likely to ever report evidence of bullet tumbling as long as the minimum twist rate is used.

However, it is common that professionals and recreational shooters desire a way to estimate bullet stability when shooting under varying environmental conditions, using a rifle and bullet combination outside of the designers' original intent (plinking loads, subsonic applications, heavier bullets, etc.), trying to enhance terminal performance with early yaw, or trying to improve accuracy by reducing the probability of tumbling when shooting through brush. Don Miller's original stability formula provided shooters with an empirical twist rule that was more accurate than the Greenhill formula. Consequently, in spite of the assumptions of constant bullet density, the twist rule was incorporated into several ballistics calculators and soon came into wide use. Later, a modified twist formula was published to more accurately predict the gyroscopic stability of plastic-tipped bullets. (Courtney and Miller, 2012a). Because of the assumption of constant bullet density, the original twist rule underestimated the stability of plastic-tipped bullets.

Collaborating with Don from summer 2010 through the fall of 2011 in the development and testing of the twist formula for plastic-tipped bullets was a tremendous experience. After the manuscripts (Courtney and Miller, 2012a; Courtney and Miller, 2012b) were delivered to Precision Shooting, Don began discussing a range of fascinating experiments.

Don was eager to debunk hypothetical concerns of "overstabilized" bullets decreasing accuracy. MC was concerned that confounding effects (barrel harmonics, varying tip off, wind, shooter error, etc.) would make those experiments challenging and tried to shift the conversation to experiments that were possible using chronographs to measure velocity, because excellent accuracy is available in velocity measurements without the factors that might tend to confound and cast doubt on accuracy based experiments.

# Aerodynamic Drag and Gyroscopic Stability

Don suggested a three chronograph arrangement to simultaneously measure the drag over near and far 50 yard intervals. We have since employed this arrangement to document pitch and yaw early in flight, the damping of pitch and yaw, and the increased drag due to pitch and yaw early in flight (Courtney et al., 2012).

Don Miller was also concerned that his original twist rule would underestimate stability of open tipped match rifle bullets that have a significant air volume in the nose. The original Miller Twist Rule was developed with the assumption of constant density throughout the bullet volume. It works relatively well for many jacketed lead bullets because the densities of lead and copper are relatively close. However, the air in open-tipped match rifle bullets introduces a significant variation from the assumption of constant density, and it causes the center of gravity to be further back than in solid-tipped designs.

One of the goals of the present study is to better document the relationship between aerodynamic drag and gyroscopic stability, because this relationship will be an important tool in development and testing of a future twist formula to more accurately estimate the stability of open tipped match bullets with a significant volume of empty space (air) in the nose. The other goal of the present study is to further test the existing twist rules by studying the relationship between aerodynamic drag and gyroscopic stability.

Care should be exercised in distinguishing gyroscopic stability from dynamic stability. Gyroscopic stability governs how a bullet flies and whether it remains point forward in early flight (muzzle to a few hundred yards or so). Bullets such as the 168 grain SMK and the M855 are known to encounter dynamic instability problems at long range even though they have adequate gyroscopic stability. A discussion of dynamic stability is beyond the scope of the current paper, but one cannot reasonably infer anything about long range dynamic stability from gyroscopic stability.

**Method**

The experimental design to measure ballistic coefficients uses two CED Millenium chronographs with LED sky screens. The experimental care employed to ensure that the optical chronographs meet their specified accuracy level (0.3%) includes ensuring the skyscreen planes are parallel, adding mechanical rigidity to ensure minimal motion during the experiment (see Figure 1), and keeping the projectile paths centered over the optical sensors in a square window 50mm x 50mm.

The chronographs are further calibrated by placing them back to back, with minimal separation, and shooting though them. Each reading of the second chronograph is adjusted upward appropriately for the small loss of velocity (< 5 fps) over the two to four foot distance from the closest chronograph. Then the average velocity of ten shots can be compared to determine systematic variations in the readings between the chronographs. In this manner, the variations between chronographs can be reduced to 0.1%.

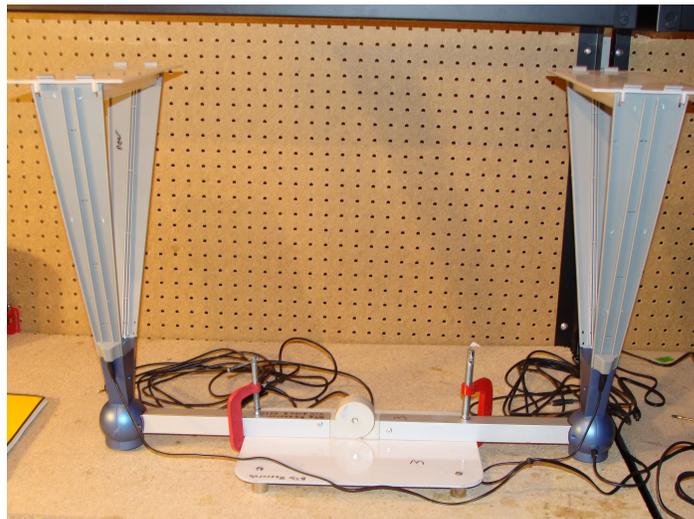

*Figure 1: CED model of optical chronograph used for velocity measurements. Note addition of mechanical rigidity by clamping chronograph rail to steel plate to reduce relative motion of skyscreens and ensure they remain parallel during measurements.*

After calibration, the chronographs are placed either 10 feet and 160 feet or 10 feet and 310 feet from the muzzle. Chronograph separations are measured with a tape measure and are accurate within a few inches. With this arrangement, velocity losses are determined over the separation interval. The experiment with the 60 grain VMAX used a 300 ft separation between the chronographs. After we learned that a 50 yard interval is more sensitive to drag increases due to bullet coning (Courtney et al., 2012), we shortened the chronograph separation to 150 feet for subsequent experiments with the 55 grain Sierra

# Aerodynamic Drag and Gyroscopic Stability

BlitzKing and the 53 grain Barnes TSX. (The Sierra Reloading Manual, 5th edition, also suggests a 50 yard interval is most appropriate for measuring BCs sensitive to gyroscopic stability and coning motions.)

Environmental conditions were measured with a Kestrel 4500 pocket weather meter. Gyroscopic stability was computed with the Courtney-Miller twist formula for plastic tipped bullets (Courtney and Miller, 2012a) and with the original Miller Twist Rule for the solid copper Barnes TSX (Miller, 2005; Miller, 2009).

Some have expressed doubt regarding the accuracy and repeatability of drag measurements using inexpensive optical chronographs. However, the accuracy of this experimental method for determining ballistic coefficients has been validated at 1% using an independent acoustic method to determine time of flight. The repeatability has been validated by measuring the ballistic coefficient of the same laboratory standard bullet on multiple occasions at a level of 1%. Further, other authors (Litz, 2009a) have also found the same model of chronograph used here to be adequate for BC measurements accurate to 1%.

As described previously (Courtney and Miller 2012b), one can shoot bullets out of the same rifle barrel over a range of stabilities by varying the velocity with different powder charges. In .223 Remington it is convenient to use two near full power loads with Varget or CFE 223 followed by a sequence of reduced loads beginning with 14 grains of Blue Dot and lowering the powder charge in 1 grain steps until a velocity is reached where the Sg is close to 1.0. After the far chronograph was hit with a tumbling bullet (Courtney and Miller, 2012b) in an earlier experiment, greater care was taken not to go far below Sg = 1.0. In .222 Remington, it is convenient to use two near full power loads with Varget or CFE 223 followed by a sequence of reduced loads beginning with 12 grains of Blue Dot and lowering the powder charge in 1 grain steps until a velocity is reached where the Sg is close to 1.0.

The approach of varying Sg by varying velocity in the same barrel is preferred to shooting from barrels with different twist rates, because other work has shown that effects from different barrels other than gyroscopic bullet stability can cause significant differences in aerodynamic drag (Litz, 2009b). Varying the velocity also allows a number of Sg values to be achieved over a range without needing a separate rifle barrel for each Sg. The possible confounding effect of BC variation with muzzle velocity is relatively small compared with the decrease of BC with Sg. For example, the BC of the 55 grain Sierra BlitzKing (SBK) is reduced from 0.271 to 0.174 as the Sg is lowered. Sierra has only measured the BC to decrease from 0.271 to 0.224 over as velocity is decreased and stability is sufficient. Similarly, reviewing BC reductions as velocity is decreased from 2500 fps to 1500 fps for .224" diameter bullets in Litz (2009a) shows 3-17% reduction in measured BCs. Therefore, it is reasonable to attribute further reductions in BC (increased drag) to lower Sg.

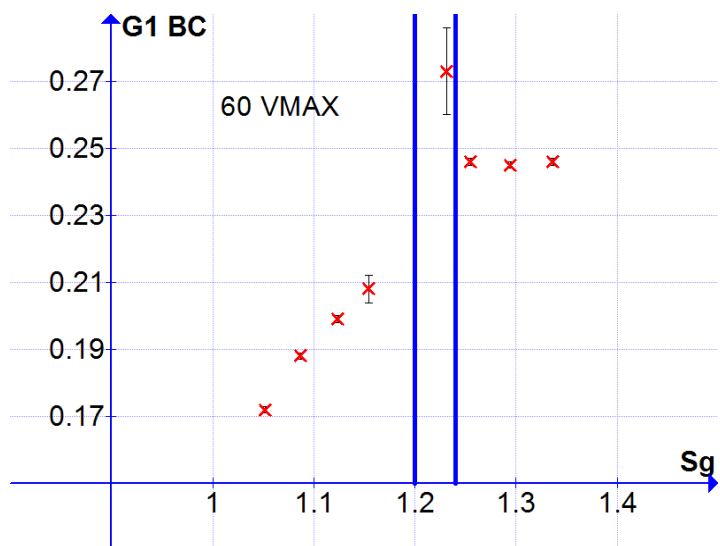

Figure 2: Measured G1 ballistic coefficient vs. gyroscopic stability of the 60 grain Hornady VMAX in .223 Remington. Note the relatively constant BC above Sg = 1.25, the significant increase in BC at Sg = 1.231, and the decreasing BC as Sg is lowered from 1.2 to 1.0.

## Results

Figure 2 shows a graph of G1 ballistic coefficient vs. gyroscopic bullet stability for the 60 grain VMAX. Ballistic coefficients are calculated by entering the near and far velocities, along with the environmental conditions (ambient pressure, humidity, and temperature) into the JBM ballistic calculator. Note the relatively constant BC for Sg above 1.25 and the decreasing BC as Sg is lowered from 1.2 toward 1.0. This is expected and is consistent with the descriptions in the 4th and 5th editions of the Sierra reloading manuals. However, also note the increase in BC at Sg = 1.231. This was unexpected. Given the relatively large error bars, we were reluctant to make much of this observation until it could be repeated for other bullets. The large error bars might be due to significant shot to shot variations in BC. However, they could indicate an inherent variability at that stability, or they could simply be attributable to some factor that gives greater variations in the tip off rate.

It is also possible to compute and plot drag coefficient (Cd) vs. gyroscopic stability, but there is a significant increase in Cd as the velocity is lowered from nearly Mach 3 down to Mach 1.2 or so which corresponds to the lower gyroscopic stabilities. Graphing BC vs. Sg reduces the impact of the transonic drag rise. Transonic drag rise is built into the G1 BC drag model so that most of the decrease in BC shown in Figure 2 is due to the decrease in gyroscopic stability rather than transonic drag rise.

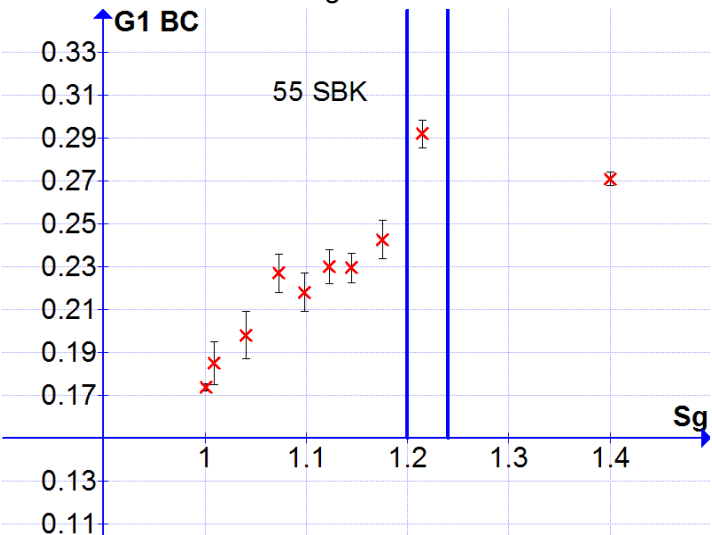

*Figure 3: Ballistic coefficient vs. gyroscopic stability of the 55 grain Sierra BlitzKing in .222 Remington. Note the significant increase in BC at Sg = 1.215, and the decreasing trend in BC as Sg is lowered from 1.2 to 1.0.*

Figure 3 shows a graph of ballistic coefficient vs. gyroscopic stability for the 55 grain Sierra BlitzKing (SBK) bullet fired from a .222 Remington with a 1 in 14" twist. As in the case of the 60 grain VMAX, there is an increase of the measured ballistic coefficient at Sg close to 1.2 (Sg = 1.215 in this case), suggesting that there is something of a "sweet spot" in stability as the stability is lowered where the bullet experiences less aerodynamic drag. As the stability is lowered from 1.2 to 1.0, the ballistic coefficient decreases significantly.

Figure 4 shows the ballistic coefficient vs. gyroscopic stability for the 53 grain Barnes TSX solid copper bullet fired from a .223 Remington with 1 in 12" twist. Note the significant increase in ballistic coefficient at Sg = 1.248, and that this increase seems to be located in a relatively narrow band because the points Sg = 1.212 and Sg = 1.268 do not display this significant decrease in drag. Note also the relatively large error bars, much like the BC increase at Sg = 1.231 in the case of the 60 grain VMAX. This suggests that even though the average drag is significantly reduced, there are much larger shot to shot variations in aerodynamic drag, probably because some shots experience rapid damping of pitch and yaw (lower drag); whereas, pitch and yaw damping may be slower in other shots. As in the case of other bullets reported here, the ballistic coefficient shows a downward trend as the stability is lowered from 1.2 to 1.0. The ten shots at the lowest predicted stability (Sg = 0.994) did not tumble. The accuracy of the twist formula is expected to be within 5%, so tumbling is expected to most likely occur somewhere between Sg = 0.95 and Sg = 1.05 and not exactly at Sg = 1.00 (Courtney and Miller 2012b).

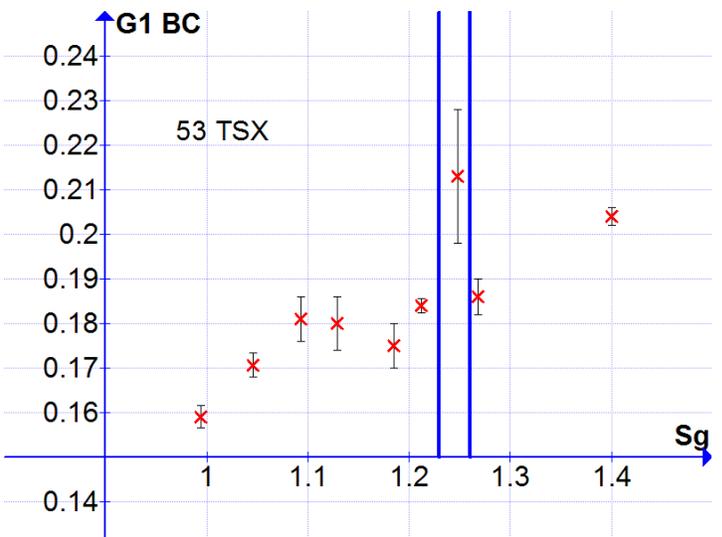

*Figure 4: Ballistic coefficient vs. gyroscopic stability of the 53 grain solid copper Barnes TSX in .223 Remington. Note the significant increase in BC at Sg = 1.248, and the decreasing trend in BC as Sg is lowered from 1.2 to 1.0.*

### Discussion

The decrease of ballistic coefficient as stability is lowered toward Sg = 1.00 is expected and has been noted in the Sierra reloading manuals, by Brian Litz (2009a), and by Courtney and Miller (2012a). The fact that it is observed here at predicted stabilities from 1.2 to 1.0 lends confidence to the suggestion of 5% accuracy in the twist formulas. The use of measured drag and observations of a reduction in BC to identify the onset of instability is on firmer experimental footing now that it has been experimentally observed in a wider array of bullets than before. This technique is probably quantitatively and experimentally more accessible than the use of yaw cards and it simultaneously quantifies the expected increase in drag during the onset of instability for a given bullet. It also further demonstrates that marginal stability likely has a negative impact on long range accuracy and bullet performance in cases where the shooter might not observe overt tumbling.

The feature of an *increase* in BC near Sg = 1.23 was somewhat unexpected. However, since it seems to be observed with some reliability, it may prove useful in further work for quantifying stability for bullets where there is not yet an accurate twist rule and for development of additional twist formulas to handle cases such as open tipped match rifle bullets with an empty volume of air in the nose.

Predicting stability with empirical twist formulas has never been an exact science. Programs like PRODAS compute the transverse and longitudinal moments of inertia to a high degree of accuracy, and thus they can be expected to make reliable predictions within the tolerances of the ability to determine bullet dimensions, density, and environmental conditions. Given the variety of bullet shapes and styles, it is unlikely that empirical twist formulas will ever match the accuracy of modeling approaches that execute complex three dimensional volume integrals. However, within their realm of applicability, it does seem that empirical twist formulas are capable of predicting stability within 5%. This means that shooters can have a high degree of confidence applying these twist formulas if due care is taken measuring bullet dimensions, rifle twist rates, and environmental conditions. Since the onset of BC variations seems to be consistently at or below Sg = 1.25, keeping the predicted stability above Sg = 1.30 is probably ample margin for error.

One might be tempted to try and design a rifle and bullet combination to realize the increased BC of the sweet spot observed near Sg = 1.23. The authors think this is a bad plan and unworkable approach. Even if the muzzle velocity can be held sufficiently constant, Sg will change with environmental conditions and shift the Sg out of the sweet spot. In addition, the larger shot to shot variations in drag will negatively impact long range accuracy and more than erase the marginal gains from an average increase in ballistic coefficient.

### Acknowledgements
This work was funded by BTG Research (www.btgresearch.org) and the United States Air Force Academy. The authors are grateful to Don Miller for his encouragement on the project and to Colorado Rifle Club and Louisiana Shooters Unlimited where the experiments were performed. This paper is dedicated to the memory of Don Miller. We are grateful for his contributions to ballistics in general and to stability science. We wish we had more time to get to know him better. The authors appreciate valuable feedback from several reviewers which has been incorporated into revisions of the manuscript.